\documentclass[showpacs]{revtex4} 
\usepackage{epsfig,amsmath,amssymb,graphicx}
\usepackage{natbib}
\usepackage{bm}
\usepackage[english]{babel} %

\begin{document}

\vspace{0mm}
\title{ ON THE RELATIVISTIC DESCRIPTION OF CHARGED FERMIONS} %
\author{Yu.M. Poluektov}
\email{yuripoluektov@kipt.kharkov.ua (y.poluekt52@gmail.com)} %
\affiliation{National Science Center ``Kharkov Institute of Physics
and Technology'',  61108
Kharkov, Ukraine}

\begin{abstract}

A method for describing charged relativistic Fermi fields is proposed, in which particles of opposite charges are treated equally and states with negative energy are excluded. The concept of charge quantum number is introduced. Fields of particles and antiparticles with different charge quantum numbers are associated with wave functions for which the Born interpretation as probability amplitudes is valid. 
\newline%
\end{abstract}
\maketitle 

\section{Introduction}\vspace{0mm} 
The relativistic equation for an electron was obtained by Dirac in his classical works \cite{Dirac}. The exposition of Dirac's theory is available in a large number of well-known textbooks and monographs \cite {AB,BSh1,BSh2,Schweber,BD1,BD2,BLP,Feynman,Weinberg,PS,Bilenky,Thirring,STZB}. Like the nonrelativistic Schrödinger equation \cite{LL}, which describes the spatial and temporal evolution of a complex function, the Dirac equation also describes a complex function that includes four components. Solutions of the nonrelativistic Schrödinger equation for a free particle with positive energy form a complete set of states, into which an arbitrary solution can be expanded. In Dirac's theory, along with solutions with positive energy, there are solutions with the opposite sign of energy. Unlike nonrelativistic theory, solutions of the Dirac equation corresponding to states with positive energy do not form a complete set of states, and to obtain a general solution, solutions with negative energy must also be considered. For the physical interpretation of states with negative energy, the "hole theory" is often used \cite{Dirac,BD1,BD2}. 
In \cite{Pl}, the author proposed a formulation of Dirac's theory in which the electron and positron are treated equally and states with negative energies are absent. The wave functions of the electron and positron admit a Born probabilistic interpretation and are considered as probability amplitudes. The aim of this paper is to further develop the approach proposed in \cite{Pl} by introducing the concept of a charge quantum number, which takes values of plus or minus one for particles of opposite charge signs. It is assumed that the complex field of a charged Fermi particle is characterized by such a charge quantum number, which describes the electron and positron states in an external electromagnetic field. In contrast to the work \cite{Pl}, where the Euclidean metric with an imaginary fourth component of 4-vectors adopted in the book \cite{AB} was used, in this work the theory is formulated using the Feynman pseudo-Euclidean metric with a real fourth component of 4-vectors. 
\section{Dirac equation for fermions with charge quantum number}\vspace{-0mm}
The Dirac equation describes the evolution of the four-component
complex function  \(\psi\left( x \right)\equiv \psi_{j}(\text{x},t) \)  where the takes values \(j=1,2,3,4 \). We will use a pseudo-Euclidean metric, which is more convenient since it does not contain an imaginary component. We assume that the Greek indices range over the values \(\mu=(0,1,2,3) \), and the Latin indices \(k=(1,2,3) \). The four-dimensional contravariant and covariant coordinates are defined by the formulas \(x^{\mu}=(x^{0}\equiv ct,\textbf{x}) \), \(x_{\mu}=(x_{0}\equiv ct,-\textbf{x}) \), \(c\) is the speed of light. We choose the 4x4 gamma matrices in the form 
 \begin{equation}\label{01}
\begin{array}{c}\displaystyle  \gamma_{0}=\gamma^{0}=\begin{pmatrix}
1&0  \\
 0&-1 
\end{pmatrix},\ \qquad \gamma^{k}=-\gamma_{k}=\begin{pmatrix}
0 &\sigma_{k}  \\
 -\sigma_{k}&0 
\end{pmatrix}, \qquad (k=1,2,3),\vspace {5mm}\\\displaystyle 
\gamma^{5}=\gamma_{5}=i\gamma^{0}\gamma^{1}\gamma^{2}\gamma^{3}=\begin{pmatrix}
0 &1  \\
 1&0 
\end{pmatrix}, \end{array}
\end{equation}
where \(\sigma_{k}\) are the Pauli matrices. In matrix notation, in the absence of an electromagnetic field, the Dirac equations for the function \(\psi(x) \) and the conjugate function \(\overline{\psi}(x)=\psi^{+}(x)\gamma^{0} \) have the form 
\begin{equation}\label{02}\displaystyle {i\gamma^{\mu}\frac{\partial \psi(x)}{\partial x^{\mu}}-\mu\psi(x)=0,}\qquad {\\i\frac{\partial \overline{\psi}(x)}{\partial x^{\mu}}\gamma^{\mu}+\mu\overline{\psi}(x)=0},
 \end{equation}
 where \(\mu=mc/\hbar \) is the inverse Compton length, \(m\) is the fermion mass and $\hbar$ is the Planck constant.
The sign \(+\) in \(\psi^{+}(x)\) denotes Hermitian conjugation, and the repeating Greek indices imply summation from 0 to 3 throughout. We will also use the metric tensor \(g^{\mu\nu}=g_{\mu\nu}\), for which \(g^{\mu\nu}=0\) if \(\mu\neq \nu\), \(g^{00}=1\) and \(g^{kk}=-1\). 

Taking into account the interaction of a fermion with a charge \(e\)  with an electromagnetic field is usually carried out using the well-known replacement  
\begin{equation}\label{03}\displaystyle \frac{\partial\psi }{\partial x^{\mu}}\to\left( \frac{\partial }{\partial x^{\mu}}+\frac{ie}{\hbar c}A_{\mu}(x) \right)\psi.
\end{equation}  
Here the 4-vector potential is introduced \(A^{\mu}=(A^{0},\textbf{A}) \), \(A_{\mu}=(A_{0},-\textbf{A}) \). Typically, the electron's charge in the Dirac equation is assumed to be negative, which introduces some asymmetry into the theory. Since particles with opposite charges must be treated equally in the theory, it is natural to introduce a quantum number \(\sigma=\pm 1\)   that determines the charge state of the field. In this case, the interaction with the electric field will be included through the replacement 
\begin{equation}\label{04}\displaystyle \frac{\partial\psi }{\partial x^{\mu}}\to\left( \frac{\partial }{\partial x^{\mu}} +\sigma\frac{i\left| e \right|}{\hbar c}A_{\mu}(x)\right)\psi. 
\end{equation}
When  \(\sigma=-1\) formula (\ref{04}) describes the interaction of electrons with the electromagnetic field, and when \(\sigma=+1\)  of positrons with the electromagnetic field. For brevity, we will use the notation \(\tilde{e}\equiv \left| e \right|/\hbar c = 1.52\cdot 10^{7}cm^{-3/2}g^{1/2}s\). Note that \(\left| e \right| \cdot \tilde{e}=\alpha=7.297\cdot 10^{-3}\)  is fine structure constant. We will also distinguish between electron and positron fields using the charge quantum index \(\psi_{\sigma}\), so that \(\psi_{-1}\equiv \psi_{e}\), \(\psi_{+1}\equiv \psi_{p}\). Taking into account the introduced notations, the Dirac equations for \(\psi_{\sigma}\left( x \right)\) and \(\overline{\psi}_{\sigma}\left( x \right)\) in an electromagnetic field will take the form 
\begin{equation}\label{05}\displaystyle i\gamma^{\mu}\left( \frac{\partial }{\partial x^{\mu}}+i\sigma\tilde{e}A_{\mu}(x) \right)\psi_{\sigma}(x)-\mu\psi_{\sigma}(x)=0,
\end{equation}
\begin{equation}\label{06}\displaystyle i\left( \frac{\partial }{\partial x^{\mu}}-i\sigma\tilde{e}A_{\mu}(x) \right)\overline{\psi}_{\sigma}(x)\gamma^{\mu}+\mu\overline{\psi}_{\sigma}(x)=0.
\end{equation}
Let \(\psi_{\sigma}\left( x \right)\)   is a solution of the Dirac equation with charge quantum number \(\sigma\) , then, as can be seen, the function 
\begin{equation}\label{07}\displaystyle  \psi_{\overline{\sigma}}(x)=C^{\ast } \widetilde{\overline{\psi}}_{\sigma}(x)
\end{equation} 
is a solution of the Dirac equation with quantum number  \(\overline{\sigma}\equiv -\sigma \). 
In (\ref{07}) \(C\)  is a unitary charge conjugation matrix that satisfies the conditions \cite{AB}: 
\begin{equation}\label{08}\displaystyle  C^{+}C=CC^{+}=1, \qquad  C\gamma_{\mu}C^{+}=-\widetilde{\gamma}_{\mu},\qquad  C=-\widetilde{C}.
\end{equation} 
A matrix satisfying these conditions is defined up to a phase factor \(e^{i\theta} \) . In particular, it can be chosen in the form \(C=\gamma^{2}\gamma^{0} \) . The symbols \(\ast\)  and \(\sim\) in (\ref{07}) and (\ref{08}) denote the operations of complex conjugation and transposition. Thus, the electron and positron fields are not independent, but are related by relation (\ref{07}). 

Let us expand the solution of the Dirac equation with a charge number  \(\sigma\) into a Fourier integral: 
\begin{equation}\label{9}\displaystyle \psi_{\sigma}(x)=\int_{-\infty }^{+\infty }b_{\sigma}(\omega)\psi_{\sigma}(\textbf{x},\omega)e^{-i\omega t}d\omega=\psi_{\sigma}^{(+)}(x)+\psi_{\sigma}^{(-)}(x),
\end{equation}
 where the positive-frequency and negative-frequency functions are defined by the formulas:
\begin{equation}\label{10}\displaystyle \psi_{\sigma}^{(\pm )}(x)=\int_{0}^{\infty }b_{\sigma}(\pm\omega)\psi_{\sigma}(\textbf{x},\pm \omega)e^{\mp i\omega t}d\omega. 
\end{equation} 
It is assumed that there exists the integral
\begin{equation} \label{11}
\begin{array}{l}
\displaystyle{%
  \frac{1}{2\pi}\int_{-\infty}^{\infty}\psi(x)e^{i\omega t}dt< \infty. %
}%
\end{array}
\end{equation}

Note that since the integration in formulas (\ref{10}) is carried out only over positive frequencies, the division of functions into positive-frequency and negative-frequency is conditional. We also have 
\begin{equation}\label{12}\displaystyle \widetilde{\overline{\psi}}_{\sigma}(x)=\int_{-\infty }^{+\infty }\widetilde{\overline{\psi}}_{\sigma}(\textbf{x},\omega)e^{i\omega t}d\omega=\widetilde{\overline{\psi}}_{\sigma}^{(+)}(x)+\widetilde{\overline{\psi}}_{\sigma}^{(-)}(x), 
\end{equation} 
where
\begin{equation}\label{13} \displaystyle \widetilde{\overline{\psi}}_{\sigma}^{(\pm )}(x)=\int_{0}^{\infty }b_{\sigma}^{\ast }(\pm \omega)\widetilde{\overline{\psi}}_{\sigma}(\textbf{x},\pm \omega)e^{\pm i\omega t}d\omega. 
\end{equation}
From (\ref{07}) follow the relationships between positive-frequency and negative-frequency functions 
\begin{equation} \label{14}\displaystyle  b_{\overline{\sigma}}(\omega)\psi_{\overline{\sigma}}(\textbf{x},\omega)=C^{\ast }b_{\sigma}^{\ast }(-\omega)\widetilde{\overline{\psi}}_{\sigma}(\textbf{x},-\omega)\end{equation}
and also
\begin{equation}\label{15}\displaystyle \psi_{\sigma}^{(+ )}(x)=C^{\ast } \widetilde{\overline{\psi}}_{\overline{\sigma}}^{(-)}(x),\quad\psi_{\sigma}^{(-)}(x)=C^{\ast }\widetilde{\overline{\psi}}_{\overline{\sigma}}^{(+)}(x).
 \end{equation}
Thus, the general solution of the Dirac equation can be expressed either through positive-frequency \(\psi_{\sigma}^{(+)}(x)\) or through negative-frequency functions \(\psi_{\sigma}^{(-)}(x)\)  of fields of different charge indices: 
\begin{equation} \label{16} \displaystyle \psi_{\sigma}(x)=\psi_{\sigma}^{(+)}(x)+C^{\ast }\widetilde{\overline{\psi}}_{\overline{\sigma}}^{(+)}(x) = \psi_{\sigma}^{(-)}(x)+C^{\ast }\widetilde{\overline{\psi}}_{\overline{\sigma}}^{(-)}(x).
\end{equation}
The general solution of the Dirac equation (\ref{16}) is expressed in terms of both the particle wave function and the antiparticle wave function, but it is not a linear superposition of these functions, since it contains an antilinear transformation of complex conjugation, and therefore it cannot be given the meaning of a probability amplitude. Since the integration in (\ref{10}) and (\ref{13}) is carried out over positive frequencies, it does not matter which form of notation is chosen in (\ref{16}). In what follows, we will always use positive-frequency functions as independent fields, omitting the sign in the notation at the top for brevity, so that in what follows we assume 
\begin{equation} \label{17} 
\begin{array} {c}\displaystyle 
\displaystyle \psi_{\sigma}^{(+)}(x)\equiv \psi_{\sigma}(x)=\int_{0}^{\infty }b_{\sigma}(\omega)\psi_{\sigma}(\textbf{x},\omega)e^{-i\omega t}d\omega,\vspace {2mm} \\\displaystyle
\widetilde{\overline{\psi}}_{\sigma}^{(+)}(x)\equiv\widetilde{\overline{\psi}}_{\sigma}(x)\equiv\underline{\psi}_{\sigma}(x)=\widetilde{\gamma}^{0}\psi_{\sigma}^{\ast }(x)=\int_{0}^{\infty }b_{\sigma}^{\ast }(\omega)\widetilde{\overline{\psi}} _{\sigma}(\textbf{x},\omega)e^{i\omega t}d\omega.
\end{array}
\end{equation}
 
These functions describe states with positive energies and opposite charges \(\sigma=\mp 1\) . It will be shown below that in a stationary field, each of the functions \(\psi_{\sigma}(x)\)  and \(\underline{\psi}_{\overline{\sigma}}(x)=\widetilde{\overline{\psi}}_{\overline{\sigma}}(x)\)  independently satisfies the Dirac equation and they can be interpreted as probability amplitudes. Therefore, these functions must be normalized by the conditions:
\begin{equation}\label{18}\displaystyle \int_{}^{}\left| \psi_{\sigma}(x) \right|^{2}d\textbf{x}=1,\quad\int_{}^{}\left| \underline{\psi}_{\overline{\sigma}}(x) \right|^{2}d\textbf{x}=1.  \end{equation}
For stationary fields, the theory can be constructed using either the function \(\psi_{\sigma}(x)\)  or the function \(\underline{\psi}_{\sigma}(x)=\widetilde{\overline{\psi}}_{\sigma}(x)\). 
\section{The Dirac equations for probability amplitudes}\vspace{-0mm} %
We substitute the general solution (\ref{16}) into equations (\ref{05}), (\ref{06}) and introduce the notation 
\begin{equation}\label{19} \displaystyle Q_{\sigma}(x)\equiv i\gamma^{\mu}\left( \frac{\partial }{\partial x^{\mu}}+i\sigma\tilde{e}A_{\mu}(x) \right)\psi_{\sigma}(x)-\mu\psi_{\sigma}(x). 
\end{equation}
Then the equation obtained as a result of such a substitution can be written in the form 
\begin{equation} \label{20} \displaystyle  Q_{\sigma}(x)+C^{\ast }\widetilde{\overline{Q}}_{\overline{\sigma}}(x)=0, 
\end{equation}
where \(\displaystyle \widetilde{\overline{Q}}_{\overline{\sigma}}(x)=\widetilde{\gamma}^{0}Q_{\overline{\sigma}}^{\ast }(x)=\widetilde{Q^{+}_{\overline{\sigma}}(x)\gamma^{0}} \), so that 

\begin{equation} \label{21} \displaystyle\widetilde{\overline{Q}}_{\overline{\sigma}}(x)\equiv-i\widetilde{\gamma}^{\mu}\left( \frac{\partial }{\partial x^{\mu}}+i\sigma\widetilde{e}A_{\mu}(x) \right)\underline{\psi}_{\overline{\sigma}}(x)-\mu\underline{\psi}_{\overline{\sigma}}(x). 
\end{equation}
Let us first consider the case of a stationary electromagnetic field, assuming  \(A_{\mu}(x)=A_{\mu}(\textbf{x})\) . Using (\ref{10}), we find 
\begin{equation} \label{22} \displaystyle Q_{\sigma}(x)=\int_{0}^{\infty }b_{\sigma}(\omega)Q_{\sigma}(\textbf{x},\omega)e^{-i\omega t}d\omega,
\end{equation}
where 
\begin{equation}\label{23} \displaystyle Q_{\sigma}(\textbf{x},\omega)=\left[ \frac{\omega}{c}\gamma^{0}+i\gamma^{k} \frac{\partial }{\partial x^{k}}-\sigma\widetilde{e}\gamma^{\mu}A_{\mu}(\textbf{x})-\mu\right]\psi_{\sigma}(\textbf{x},\omega).
\end{equation}
From (\ref{20}) and (\ref{22}) we have 
\begin{equation} \label{24} \displaystyle \int_{0}^{\infty }\left[ b_{\sigma}(\omega)Q_{\sigma}(\textbf{x},\omega)e^{-i\omega t}+b_{\overline{\sigma}}^{\ast }(\omega)C^{\ast } \widetilde{\overline{Q}}_{\overline{\sigma}}(\textbf{x},\omega)e^{i\omega t}\right]d\omega=0.  
\end{equation}
Multiplying (\ref{24}) first by \(e^{i\omega't}\) , where  \({\omega'} \ge {0}\),  and integrating over time, and then multiplying by \(e^{-i\omega't}\), and also integrating over time, we obtain that \(Q_{\sigma}(\textbf{x},\omega)\)  and  \(\widetilde{\overline{Q}}_{\overline{\sigma}}(\textbf{x},\omega)\). Thus, we arrive at independent equations for fields with opposite charge quantum numbers and positive frequency: 
\begin{equation} \label{25} \displaystyle \left[ i\gamma^{k}\frac{\partial }{\partial x^{k}}+\frac{\omega}{c}\gamma^{0}-\sigma\widetilde{e}\gamma^{\mu}A_{\mu}(\textbf{x})-\mu \right]\psi_{\sigma}(\textbf{x},\omega)=0, 
\end{equation}
\begin{equation} \label{26} \displaystyle \left[ i\widetilde{\gamma}^{k}\frac{\partial }{\partial x^{k}}-\frac{\omega}{c}\widetilde{\gamma}^{0}+\sigma\widetilde{e} \widetilde{\gamma}^{\mu}A_{\mu}(\textbf{x})+\mu\right]\underline{\psi}_{\sigma}(\textbf{x},\omega)=0, 
\end{equation}
where \(\underline{\psi}_{\sigma}(\textbf{x},\omega)\equiv \widetilde{\overline{\psi}}_{\sigma}(\textbf{x},\omega)\). 
For the time-dependent wave functions \(\psi_{\sigma}(x)\), \(\underline{\psi}_{\overline{\sigma}}(x)\)   (\ref{17}) and their conjugate functions, the equations have the form 
\begin{equation}\label{27} \displaystyle i\gamma^{\mu}\frac{\partial\psi_{\sigma}(x) }{\partial x^{\mu}}-\left[ \sigma\widetilde{e}\gamma^{\mu}A_{\mu}(x)+\mu\right]\psi_{\sigma}(x)=0, 
\end{equation} 
\begin{equation} \label{28}\displaystyle i\frac{\partial\overline{\psi}_{\sigma}(x) }{\partial x^{\mu}}\gamma^{\mu}+\overline{\psi}_{\sigma}(x)\left[ \sigma\widetilde{e}\gamma^{\mu}A_{\mu}(x)+\mu \right]=0, 
\end{equation}
\begin{equation} \label{29} \displaystyle i\widetilde{\gamma}^{\mu}\frac{\partial \underline{\psi}_{\overline{\sigma}}(x)}{\partial x^{\mu}}-\left[ \sigma\widetilde{e}A_{\mu}(x)\widetilde{\gamma}^{\mu}-\mu \right]\underline{\psi}_{\overline{\sigma}}(x)=0,
\end{equation}
\begin{equation} \label{30} \displaystyle i\frac{\partial \underline{\overline{\psi}}_{\overline{\sigma}}(x)}{\partial x^{\mu}}\widetilde{\gamma}^{\mu}+\overline{\underline{\psi}}_{\overline{\sigma}}(x)\left[ \sigma\widetilde{e}A_{\mu}(x)\widetilde{\gamma}^{\mu}-\mu \right]=0. 
\end{equation}
Here \(\overline{\underline{\psi}}_{\overline{\sigma}}(x)\equiv \underline{\psi}^{+}_{\overline{\sigma}}(x)\widetilde{\gamma}_{0}=\widetilde{\psi}_{\overline{\sigma}}(x)\) . In what follows, with the exception of the section on the non-stationary field, we will use the equations for the function \(\psi_{\sigma}(x)\). Note also that the Dirac equation can be written in Hamiltonian form 
\begin{equation}\label{31} \displaystyle i\frac{\partial \psi_{\sigma}}{\partial x}=\text{H}\psi_{\sigma}, 
\end{equation}
where the Hamiltonian has the form 
\begin{equation} \label{32} \displaystyle \text{H}=\gamma^{0}\gamma^{k}\left( -i\frac{\partial }{\partial x^{k}}+\sigma\widetilde{e}A_{k} \right)+\sigma\widetilde{e}A_{0}+\mu\gamma^{0}. 
\end{equation}
From the normalization conditions (\ref{18}) it follows that the relation must be satisfied 
\begin{equation}\label{33} \displaystyle 	\int_{}^{}\left| \psi_{\sigma}(x) \right|^{2}d\textbf{x}=\int_{0}^{\infty }\left| b_{\sigma}(\omega) \right|^{2}d\omega.
\end{equation}
The coefficient \(b_{\sigma}(\omega)\)  has the meaning of the probability amplitude that in a state with a wave function \(\psi_{\sigma}(x)\), a particle can be detected in the frequency range from \(\omega\) to \(\omega+d\omega\). Substituting into the left side of formula (\ref{33})  \(\displaystyle \psi_{\sigma}(x)=\int_{0}^{\infty }b_{\sigma}(\omega)\psi_{\sigma}(\textbf{x},\omega)e^{-i\omega t}d\omega\) and comparing with the right side, we find an expression for the expansion coefficient 
\begin{equation} \label{34}\displaystyle b_{\sigma}(\omega)e^{-i\omega t}=\int_{}^{}d\textbf{x}\psi^{+}_{\sigma}(\textbf{x},\omega)\psi_{\sigma}(x). 
\end{equation}
Substituting the function \(\psi_{\sigma}(x)\)  into (\ref{34}), we arrive at the orthonormality condition 
\begin{equation} \label{35} \displaystyle \int_{}^{}d\textbf{x}\psi^{+}_{\sigma}(\textbf{x},\omega)\psi_{\sigma}(\textbf{x},\omega')=\delta(\omega-\omega'). 
\end{equation}
Also, substituting (\ref{34}) into the expansion of the function \(\psi_{\sigma}(x)\) , we arrive at the completeness condition 
\begin{equation}\label{36} \displaystyle \int_{}^{}d\omega\psi_{\sigma j}^{\ast }(\textbf{x}',\omega)\psi_{\sigma k}(\textbf{x},\omega')=\delta(\textbf{x}'-\textbf{x})\delta_{jk}, 
\end{equation}
where \(j,k=1,2,3,4\)  are the components of the column \(\psi_{\sigma}(\textbf{x},\omega)\). For fields with different charges, orthogonality relations are satisfied 
\begin{equation} \label{37} \displaystyle \int_{}^{}d\textbf{x}\widetilde{\psi}_{\overline{\sigma}}(\textbf{x},\omega)C\gamma^{0}\psi_{\sigma}(\textbf{x},\omega')=0  
\end{equation}
and also
\begin{equation} \label{38} \displaystyle \int_{}^{}d\textbf{x}\widetilde{\psi}_{\overline{\sigma}}(\textbf{x},\omega)C\gamma^{0}\psi_{\sigma}(\textbf{x},\omega')=0. 
\end{equation}
Thus, the solutions of equation (\ref{25}) form a complete set \(\left\{ \psi_{\sigma}(\textbf{x},\omega) \right\}\)  in which an arbitrary solution of the Dirac equation \(\psi_{\sigma}(x)\)  with a charge quantum number \(\sigma\)  and positive energy can be expanded. 
\section{Lagrangian formalism. Conservation laws}\vspace{0mm}
We formulate the proposed approach to describing charged particles and antiparticles in terms of probability amplitudes using the Lagrangian formalism, which will allow us to obtain the energy-momentum tensor and conservation laws. The Dirac equations (\ref{27}), (\ref{28}) can be obtained if the density of the Lagrangian function is chosen in the form 
\begin{equation} \label{39}\displaystyle \Lambda_{\sigma}=\frac{i}{2}\left( \overline{\psi}_{\sigma}\gamma^{\mu}\frac{\partial\psi_{\sigma} }{\partial x^{\mu}} -\frac{\partial \overline{\psi}_{\sigma}}{\partial x^{\mu}}\gamma^{\mu}\psi_{\sigma}\right)-\sigma\widetilde{e}A_{\mu}\overline{\psi}_{\sigma}\gamma^{\mu}\psi_{\sigma}-\mu\overline{\psi}_{\sigma}\psi_{\sigma}.
\end{equation}
In what follows, where it does not cause misunderstanding, in this section we will omit the charge index \(\sigma\)  for brevity, so that further \(\Lambda\equiv \Lambda_{\sigma}\) and \(\psi\equiv \psi_{\sigma} \) . Equations (\ref{27}), (\ref{28}) follow from the Euler-Lagrange equations 
\begin{equation} \label{40}\frac{\partial \Lambda}{\partial \psi}-\frac{\partial }{\partial x^{\mu}}\frac{\partial \Lambda} { \partial \displaystyle {\frac{\partial \psi}{\partial x^{\mu}}}}=0,\qquad \frac{\partial \Lambda}{\partial \overline{\psi}}-\frac{\partial }{\partial x^{\mu}}\frac{\partial \Lambda}{ \partial \displaystyle {\frac{\partial \overline{\psi}}{ \partial x^{\mu}}}}=0. 
\end{equation}
Note that substitution of functions that are solutions of the Dirac equation into (\ref{39}) turns the Lagrangian to zero. 

From the invariance of the Lagrangian (\ref{39}) with respect to phase transformations 
\begin{equation}\label{41} \displaystyle \psi(x)\to \psi'(x)=\psi(x)e^{i\chi},\qquad\overline{\psi}(x)\to \overline{\psi}'(x)=\overline{\psi}(x)e^{-i\chi},
\end{equation}
where \(\chi\)  is a real parameter, the continuity equation for the probability density follows 
\begin{equation} \label{42} \displaystyle \frac{\partial j^{\mu}}{\partial x^{\mu}}=0.  
\end{equation}
In (\ref{42}) the 4-vector of the probability flux density has the form 
\begin{equation}\label{43} \displaystyle j^{\mu}=c\,\overline{\psi}\gamma^{\mu}\psi
\end{equation}
or \(j^{\mu}=\left( c\psi^{+}\psi,\,c\psi^{+}\gamma^{0}\gamma^{k}\psi \right)\) . Thus, the continuity equation (\ref{42}) is 
\begin{equation}\label{44} \displaystyle \frac{\partial (\psi^{+}\psi)}{\partial t}+c\frac{\partial }{\partial x^{k}}\left( \psi^{+}\gamma^{0}\gamma^{k}\psi \right)=0. 
\end{equation}
From (\ref{44}) follows the law of conservation of total probability for a particle \(\int_{}^{}d\textbf{x}\psi^{+}(x)\psi(x)=const \). 

The charge flux density vector obviously has the form 
\begin{equation} \label{45} \displaystyle j^{\mu}_{\sigma}=\sigma\left| e \right|c\overline{\psi}_{\sigma}\gamma^{\mu}\psi_{\sigma}.
\end{equation}
From the continuity equation for current density follows the law of conservation of charge \(\sigma\left| e \right|\int_{}^{}d\textbf{x}\psi_{\sigma}^{+}(x)\psi_{\sigma}(x)=\sigma\left| e \right|\). 

Let us consider the consequences of the invariance of the Lagrangian density with respect to infinitesimal coordinate transformations \(x^{\mu}\to x'^{\mu}+\delta x^{\mu}\)  when the condition is valid 
\begin{equation}\label{46}\begin{array}{c} \displaystyle \Lambda\left( \psi(x),\overline{\psi} (x),\frac{\partial \psi(x)}{\partial x^{\mu}}, \displaystyle \frac{\partial \overline{\psi}(x)}{\partial x^{\mu}},A_{\mu}(x)\right)=
\vspace {2mm}\\\displaystyle \Lambda\left( \psi'(x'),\overline{\psi'} (x'),\frac{\partial \psi'(x')}{\partial x'^{\mu}},\frac{\partial \overline{\psi'}(x')}{\partial x'^{\mu}},A'_{\mu}(x')\right), \end{array}
\end{equation}
moreover 
\begin{equation} \label{47} \displaystyle\psi(x)=\psi'(x')\quad A_{\mu}(x)=A'_{\mu}(x'). 
\end{equation}
For small \(\delta x^{\mu}\)  we have 
\begin{equation}\label{48} \displaystyle\delta \Lambda+\frac{\partial \Lambda}{\partial x^{\mu}}\delta x^{\mu}=0,
\end{equation}
where 
\begin{equation}\label{49} \begin{array}{c} \displaystyle\delta \Lambda=\Lambda\left( \psi'(x),\overline{\psi'} (x),\frac{\partial \psi'(x)}{\partial x^{\mu}},\frac{\partial \overline{\psi'}(x)}{\partial x^{\mu}},A'_{\mu}(x)\right)-\vspace {2mm}\\\displaystyle -\Lambda\left( \psi(x),\overline{\psi} (x),\frac{\partial \psi(x)}{\partial x^{\mu}},\frac{\partial \overline{\psi}(x)}{\partial x^{\mu}},A_{\mu}(x)\right).  \end{array}
\end{equation}
With this transformation, the functions also change 
\begin{equation}\label{50}\displaystyle \psi'(x)=\psi(x)+\delta \psi(x),\quad A_{\mu}'(x)=A_{\mu}(x)+\delta A_{\mu}(x),\quad \delta A_{\mu}(x)=-\frac{\partial A_{\mu}(x)}{\partial x^{\nu}} \delta x^{\nu}.\end{equation}
From formulas (\ref{48}) - (\ref{50}) follows the continuity equation 
\begin{equation}\label{51}\displaystyle \frac{\partial }{\partial x^{\mu}}\left( \frac{\partial \Lambda} {\displaystyle {\partial \frac{\partial \psi}{\partial x^{\mu}}}}\delta\psi+\delta \overline{\psi}\frac{\partial \Lambda}{\partial \displaystyle \frac{\partial \overline{\psi}}{\partial  x^{\mu}}}+\delta^{\mu}_{\nu}\Lambda \delta x^{\nu}\right)+\frac{\partial \Lambda}{\partial A_{\mu}}\delta A_{\mu}-\Lambda \delta^{\mu}_{\nu}\frac{\partial \delta x^{\nu} }{\partial x^{\mu}}=0.
\end{equation}

\textit{Small translations}.  Let us first consider the case of small translations to a constant vector \(\delta x^{\nu}=a^{\nu} \). Then from (\ref{51}) we obtain 
\begin{equation} \label{52}\displaystyle \frac{\partial T^{.\mu}_{\nu.} }{\partial x^{\mu}}=\hbar c\frac{\partial \Lambda}{\partial A_{\mu}}\frac{\partial A_{\mu}}{\partial x^{\nu}}=-\sigma\left| e \right|\overline{\psi}\gamma^{\mu}\psi\frac{\partial A_{\mu}}{\partial x^{\nu}}, 
\end{equation}
where 
\begin{equation}\label{53}\displaystyle T^{.\mu}_{\nu.}=\hbar c\left( \frac{\partial \Lambda}{\partial \displaystyle \frac{\partial \psi}{\partial x^{\mu}}}\frac{\partial \psi}{\partial x^{\nu}}+\frac{\partial \overline{\psi}}{\partial x^{\nu}}\frac{\partial \Lambda}{\partial\displaystyle  \frac{\partial \overline{\psi}}{\partial x_{\mu}}} \right)=\hbar c\frac{i}{2}\left( \overline{\psi}\gamma^{\mu}\frac{\partial \psi}{\partial x^{\nu}} -\frac{\partial \overline{\psi}}{\partial x^{\nu}}\gamma^{\mu}\psi\right)
\end{equation}
is energy-momentum tensor. Here it was taken into account that for functions satisfying the Dirac equation, as noted, the Lagrangian is equal to zero. The points in mixed tensors are introduced to indicate the order of indices. So \(T^{.\mu}_{\nu.} \)  means that \(\nu\)  is the first index, and \(\mu\)  is the second. On the right side of the continuity equation (\ref{52}) there is a source associated with the presence of an external field. Let us also present the 4-vector of energy-momentum 
\begin{equation} \label{54} \displaystyle T^{\nu}=\int_{}^{}T^{\nu 0}d\textbf{x}=i\hbar c\int_{}^{}\psi^{+}\frac{\partial \psi}{\partial x_{\nu}}d\textbf{x}.
\end{equation}
From (\ref{54}) the expressions for energy 
\begin{equation}\label{55}\displaystyle E=T^{0}=i\hbar \int_{}^{}\psi^{+}\frac{\partial\psi }{\partial t}d\textbf{x}
\end{equation} 
and momentum 
\begin{equation} \label{56}\displaystyle P^{i}=\frac{1}{c}T^{i}=-i\hbar \int_{}^{}\psi^{+}\frac{\partial \psi}{\partial x^{i}}d\textbf{x} 
\end{equation} 
follow. If we use the general solution (\ref{16}) as a wave function that has the meaning of a probability amplitude, then in this case the Dirac field energy does not have a definite sign. If, however, solutions (\ref{17}) are considered as wave functions for particles of both signs, as proposed in this approach, then the field energy for particles of both charges is always positive. Indeed, taking into account the orthonormalization condition (\ref{35}), we arrive at an obviously positive expression for the field energy with a charge quantum number \(\sigma\) : 
\begin{equation} \label{57}\displaystyle E_{\sigma}=\hbar \int_{0}^{\infty }d\omega\omega\left| b_{\sigma}(\omega)\right|^{2}. 
\end{equation}

\textit{Homogeneous Lorentz transformations.} Let us consider the consequence of the invariance of the Lagrangian density under homogeneous Lorentz transformations 
\begin{equation}\label{58}\displaystyle x'^{\mu}=L^{\mu.}_{.\nu}x^{\nu}. 
\end{equation}
It follows from \(x'^{\mu}x'_{\mu}=x^{\mu}x_{\mu}\)  condition that 
\begin{equation}\label{59}\displaystyle \underline{L}^{\rho.}_{.\mu}L^{\mu.}_{.\nu}=\delta^{\rho}_{\nu},
\end{equation}
where 
\begin{equation}\label{60}\displaystyle \underline{L}^{\rho.}_{.\mu}\equiv g_{\mu\lambda}L^{\lambda.}_{.\sigma}g^{\sigma\rho}. 
\end{equation}
Let us note that 
\begin{equation}\label{61}\displaystyle \underline{L}^{0.}_{.0}=L^{0.}_{.0},\; \; \;\underline{L}^{0.}_{.k}=-L^{k.}_{.0},\; \; \;\underline{L}^{k.}_{.0}=-L^{0.}_{.k},\; \; \;\underline{L}^{k.}_{.i}=L^{i.}_{.k}. 
\end{equation} 
For small transformations \(L^{\mu.}_{.\nu}=\delta^{\mu}_{\nu}+\epsilon^{\mu.}_{.\nu}, \;\;\; \underline{L}^{\rho.}_{.\mu}=\delta^{\rho}_{\mu}+\underline{\epsilon}^{\rho.}_{.\mu}\) , where \(\underline{\epsilon}^{\rho.}_{.\mu}\equiv g_{\mu\lambda}\epsilon^{\lambda.}_{.\sigma}g^{\sigma\rho} \) , we have 
\begin{equation} \label{62}\displaystyle x'^{\mu} =x^{\mu}+\epsilon^{\mu.}_{.\nu}x^{\nu} .
\end{equation} 
From (\ref{59}) it follows 
\begin{equation}\label{63}\displaystyle \underline{\epsilon}^{\mu.}_{.\rho}=-\epsilon^{\mu.}_{.\rho},\; \;\;\epsilon_{\mu\rho}=-\epsilon_{\rho\mu},\; \;\;\epsilon^{\mu\rho}=-\epsilon^{\rho\mu}. 
\end{equation}
When moving to a new coordinate system, the bispinor is transformed as follows:
\begin{equation}\label{64}\displaystyle \psi'(x')=S\psi(x),\quad\overline{\psi}'(x')=\overline{\psi}(x)S^{-1}, 
\end{equation}
where the matrix \(S\)  satisfies the conditions 
\begin{equation} \label{65}\displaystyle S^{-1}\gamma^{\mu}S=L^{\mu.}_{.\nu}\gamma^{\nu} 
\end{equation}
and
\begin{equation} \label{66}\displaystyle S^{+}\gamma^{0}=\gamma^{0}S^{-1}. 
\end{equation} 
With small transformations this matrix has the form 
\begin{equation}\label{67}\displaystyle  S=1+\frac{1}{2}\Sigma^{\mu\nu}\epsilon_{\nu\mu},\quad S^{-1}=1-\frac{1}{2}\Sigma^{\mu\nu}\epsilon_{\nu\mu}. 
\end{equation}
Here the quantities \(\Sigma^{\mu\nu}=-\Sigma^{\nu\mu} \)  are 4x4 matrices. From (\ref{64}), (\ref{65}) it follows that the following conditions must be satisfied 
\begin{equation} \label{68}\displaystyle \gamma^{\rho}\Sigma^{\mu\nu}-\Sigma^{\mu\nu}\gamma^{\rho}=g^{\rho\nu}\gamma^{\mu}-g^{\rho\mu}\gamma^{\nu},
\end{equation}
\begin{equation} \label{69}\displaystyle \left( \Sigma^{\mu\nu} \right)^{+}\gamma^{0}=-\gamma^{0}\Sigma^{\mu\nu}. 
\end{equation}
These conditions will be met if we assume that 
\begin{equation}\label{70}\displaystyle  \Sigma^{\mu\nu}=\frac{i}{2}\sigma^{\mu\nu},
\end{equation}
where 
\begin{equation}\label{71}\displaystyle \sigma^{\mu\nu}=\frac{i}{2}\left( \gamma^{\mu}\gamma^{\nu} -\gamma^{\nu}\gamma^{\mu}\right). 
\end{equation}
Thus, under infinitely small homogeneous Lorentz transformations we have 
\begin{equation}\label{72}\begin{array}{c} \displaystyle\delta \psi(x)=-\frac{\partial\psi(x) }{\partial x^{\mu}}x_{\nu}\epsilon^{\mu\nu}+\frac{1}{2}\Sigma_{\nu\mu}\epsilon^{\mu\nu}\psi(x), \vspace {2mm}\\
\displaystyle\delta\overline{\psi}(x)=-\frac{\partial \overline{\psi}(x)}{\partial x^{\mu}}x_{\nu}+\frac{1}{2}\overline{\psi}(x)\overline{\Sigma}_{\nu\mu}\epsilon^{\mu\nu},  \end{array}
\end{equation}
where \(\overline{\Sigma}_{\nu\mu}=\gamma^{0}\Sigma^{+}_{\nu\mu}\gamma^{0}=-\Sigma_{\nu\mu}\) . Substituting (\ref{72}) into (\ref{51}) and taking into account that \({\partial \delta x^{\nu}}/\partial{x^{\nu}}=0\) , we find 
\begin{equation}\label{73}\begin{array}{c} \displaystyle\frac{\partial }{\partial x^{\mu}}\left[ \frac{1}{c}\left( T^{.\mu}_{\rho.}x_{\nu}-T^{.\mu}_{\nu.}x_{\rho} \right)-\frac{\hbar }{4}\overline{\psi}\left( \gamma^{\mu}\sigma_{\nu\rho}+\sigma_{\nu\rho}\gamma^{\mu} \right) \psi\right]=\vspace {3mm}\\
\displaystyle=-\hbar \frac{\partial \Lambda}{\partial A_{\mu}}\left( \frac{\partial A_{\mu}}{\partial x^{\rho}}x_{\nu}- \frac{\partial A_{\mu}}{\partial x^{\nu}}x_{\rho}\right)=\hbar \sigma\widetilde{e}\overline{\psi}\gamma^{\mu}\psi\left( \frac{\partial A_{\mu}}{\partial x^{\rho}}x_{\nu}- \frac{\partial A_{\mu}}{\partial x^{\nu}}x_{\rho}\right).
 \end{array}
 \end{equation}
The density tensors of the orbital and spin moments are determined by the relations 
\begin{equation}\label{74} \displaystyle L^{\mu}_{\rho\nu}=\frac{1}{c}\left( T^{.\mu}_{\rho.}x_{\nu} -T^{.\mu}_{\nu.}x_{\rho}\right), \end{equation}
\begin{equation}\label{75}\displaystyle S^{\mu}_{\rho\nu}=-\frac{\hbar }{4}\overline{\psi}\left( \gamma^{\mu}\sigma_{\rho\nu} +\sigma_{\rho\nu}\gamma^{\mu}\right)\psi.  \end{equation}
Thus, equation (\ref{73}) expresses the law of conservation of total angular momentum 
\begin{equation}\label{76}\displaystyle M^{\mu}_{\rho\nu}=L^{\mu}_{\rho\nu}+S^{\mu}_{\rho\nu}.
\end{equation}
By integrating the components \(L^{0}_{\rho\nu}\) and \(S^{0}_{\rho\nu}\)  over 3-space, we obtain the orbital and spin moment tensors 
\begin{equation} \label{77}\displaystyle L_{\rho\nu}=\int_{}^{}d\textbf{x}L^{0}_{\nu\rho},\quad S_{\rho\nu}=\int_{}^{}d\textbf{x}S^{0}_{\nu\rho}.\end{equation} 
The axial vectors of the orbital and spin angular momentum are expressed through the spatial components of the tensors (\ref{77}): 
\begin{equation}\label{78}\displaystyle L_{i}=\frac{1}{2}\epsilon_{ikl}L_{kl}, \quad S_{i}=\frac{1}{2}\epsilon_{ikl}S_{kl}, 
\end{equation}
where  \(\epsilon_{ikl}\) is the antisymmetric unit tensor. 
\section{Free particles and antiparticles}\vspace{0mm}
In the standard formulation of Dirac's theory, when describing free particles, it is necessary to consider states with positive energy (electrons) and negative energy (positrons) separately. In this section we present the relationships for free fields in the proposed approach, when particles of both signs have positive energy and are considered on an equal basis. In the absence of an electromagnetic field, particles of different electric charges are indistinguishable. For this reason, we will omit the charge index  \(\sigma\) in the formulas of this section. The solutions of the Dirac equation (\ref{27}) without an external field for states with a certain momentum \(\hbar \textbf{k}\) have the form 
\begin{equation}\label{79}\displaystyle \psi(\textbf{x},t)=\frac{u(\textbf{k})}{\sqrt{V}}e^{i(\textbf{kx}-\omega t)}, 
\end{equation}
where the bispinor \(u(\textbf{k})\)  satisfies the equation 
\begin{equation}\label{80}\displaystyle \left(\widehat{k}-\mu\right)u(\textbf{k})=0, 
\end{equation} 
\(\left( \widehat{k}\equiv k_{\mu}\gamma^{\mu},\;\;k_{\mu}=(\omega/c,-\textbf{k}) \right)\) and can be represented as 
\begin{equation}\label{81}\displaystyle  u(\textbf{k})=\sqrt{\frac{\epsilon+\mu}{2\epsilon}}\displaystyle\begin{bmatrix}
\varphi(\textbf{k}) \\ \text{\Large$\frac{\textbf{k}\boldsymbol{\sigma}}{\epsilon+\mu}$}\varphi(\textbf{k})
\end{bmatrix}. 
\end{equation}
Since for particles of both types we consider functions (\ref{10}), for which the Fourier expansion is carried out only for positive frequencies, the root with a negative sign for energy should not be taken into account, therefore 
\begin{equation}\label{82}\displaystyle\hbar \omega=\hbar\ c\epsilon\quad \epsilon\equiv \sqrt{\mu^{2}+k^{2}}.
\end{equation}
When the normalization conditions for the spinor \(\varphi^{+}(\textbf{k})\varphi(\textbf{k})=1\)  are satisfied, the bispinor (\ref{81}) is normalized by the condition \(u^{+}(\textbf{k})u(\textbf{k})=1\). Instead of the bispinor (\ref{81}), one can use the bispinor  
\begin{equation}\label{83}\displaystyle u(k)\equiv \sqrt{\frac{\varepsilon}{\mu}}u(\textbf{k})=\sqrt{\frac{\epsilon+\mu}{2\mu}}\displaystyle\begin{bmatrix}
\varphi(\textbf{k}) \\ \text{\Large$\frac{\textbf{k}\boldsymbol{\sigma}}{\epsilon+\mu}$}\varphi(\textbf{k})
\end{bmatrix}
\end{equation}
for which the relativistically invariant normalization \(\overline{u}(k)u(k)=1 \)  is valid. In the rest system, the bispinor takes the form 
\begin{equation}\label{84}\displaystyle  u(\mu)=\displaystyle\begin{bmatrix}
\varphi(0) \\ 0 
\end{bmatrix}.
\end{equation}
In a system where the particle momentum is \(\hbar \textbf{k}\) , the bispinor (\ref{83}) can be obtained using the transformations 
\begin{equation}\label{85}\displaystyle u(k)=L(\textbf{k})u(\mu),\qquad \overline{u}(k)=\overline{u}(\mu)L^{-1}(\textbf{k}),
\end{equation}
where the matrix \(L(\textbf{k})\)  satisfies the condition 
\begin{equation}\label{86}\displaystyle L^{-1}(\textbf{k})\gamma^{\mu}L(\textbf{k})=L^{\mu.}_{.\nu}\gamma^{\nu}
\end{equation}
and has the form 
\begin{equation}\label{87}\displaystyle L(\textbf{k})=\sqrt{\frac{\epsilon+\mu}{2\mu}}\left( 1-\frac{k_{i}\gamma^{i}}{\epsilon+\mu}\gamma^{0} \right),\quad L^{-1}(\textbf{k})=\sqrt{\frac{\epsilon+\mu}{2\mu}}\left( 1+\frac{k_{i}\gamma^{i}}{\epsilon+\mu}\gamma^{0} \right).
\end{equation}
To describe the spin state of a particle, we assume that the spinor satisfies the equation 
\begin{equation} \label{88}\displaystyle \boldsymbol{\sigma}\textbf{u}\varphi^{s}(0)=s\varphi^{s}(0),  
\end{equation}
and \(\textbf{u}\) is a unit vector in the rest system. From (\ref{88})  it follows that 
\begin{equation} \label{89} s\textbf{u}=\varphi^{s+}(0)\boldsymbol{\sigma}\varphi^{s}(0). 
\end{equation}
Thus, when \(s=+1\)  the vector \(\textbf{u}\)  determines the direction of the average spin value. Formula (\ref{88}) can be written as \(\boldsymbol{\Sigma}\textbf{u}u^{s}(\mu)=su^{s}(\mu)\) , where 
\begin{equation}\label{90}\displaystyle  \Sigma_{i}=\gamma^{5}\gamma^{i}\gamma^{0}=\begin{pmatrix}
\sigma_{i} &  0\\
 0 & \sigma_{i}
\end{pmatrix}.
\end{equation}
Let us define a covariant unit 4-vector in the rest system \(u_{\mu}\equiv \left( 0,-\textbf{u} \right) \) . Then \(u_{i}\gamma^{i}=-u_{\mu}\gamma^{\mu}\)  and taking into account \(\gamma^{0}u^{s}(\mu)=u^{s}(\mu)\)  we have 
\begin{equation}\label{91}\displaystyle \gamma^{5}u_{\mu}\gamma^{\mu}u^{s}(\mu)=-su^{s}(\mu). 
\end{equation}
Using (\ref{85}), (\ref{86}), (\ref{87}) and taking into account that 
\begin{equation}\label{92}\displaystyle L^{-1}(\textbf{k})\gamma^{5}\gamma^{\mu}L(\textbf{k})={L}^{\mu.}_{.\nu}\gamma^{\nu},
\end{equation}
let us write relation (\ref{91}) in the system where the particle has momentum  \(\hbar \textbf{k}\). We have 
\begin{equation}\label{93}\displaystyle\gamma^{5}\widehat{n}u^{s}(k)=-su^{s}(k),\quad \overline{u}^{s}(k)\widehat{n}\gamma^{5}=s\overline{u}^{s}(k), \end{equation}
where \(n_{\mu}=\underline{L}^{\nu.}_{.\mu}u_{\nu} \), \(\widehat{n}\equiv n_{\mu}\gamma^{\mu} \), \(u^{s}(k)=L(\textbf{k})u^{s}(\mu) \). It is obvious that \(n_{\mu}k^{\mu}=u_{0}\mu=0 \)  and \(n_{\mu}n^{\mu}=1 \) . Thus, a bispinor \(u^{s}(k)\)  that is an eigenfunction of an operator  \(\widehat{k}\equiv k_{\mu}\gamma^{\mu} \) is also an eigenfunction of an operator \(\gamma^{5}\widehat{n} \) . Operators  \(\widehat{k} \) and \(\gamma^{5}\widehat{n} \)  commute with each other. Multiplying the first equation (\ref{93}) on the left by \(\overline{u}^{s}(k)\gamma^{\mu}\gamma^{5} \) , and the second on the right by \(u^{s}(k)\gamma^{5}\gamma^{\mu} \) , and adding them together we get 
\begin{equation} \label{94}\displaystyle \overline{u}^{s}(k)\left( \gamma^{\mu}\widehat{n}+\widehat{n}\gamma^{\mu} \right)u^{s}(k)=2s\overline{u}^{s}(k)\gamma^{5}\gamma^{\mu}u^{s}(k). 
\end{equation}
Using the commutation relations for matrices \(\gamma\), taking into account the normalization condition \(\overline{u}^{s}(k)u^{s}(k)=1\), we find 
\begin{equation}\label{95}\displaystyle \overline{u}^{s}(k)\gamma^{5}\gamma^{\mu}u^{s}(k)=sn^{\mu}.
\end{equation}
As we see, the average value of the operator \(\gamma^{5}\gamma^{\mu}\)  in the state described by the bispinor \(u^{s}(k)\)  is equal to \(sn^{\mu}\) . The relationship between the 4-vector of polarization \(n^{\mu}\)  and the unit vector \(\textbf{u}\)  is found using the Lorentz transformation:
\begin{equation}\label{96}\displaystyle \textbf{n}=\textbf{u}+\frac{\textbf{k(ku)}}{\mu(\epsilon+\mu)},\qquad n_{0}=\frac{\textbf{ku}}{\mu}. 
\end{equation}
Thus, the state of a particle of any sign with positive energy in an arbitrary reference system is completely described by its momentum and discrete coordinate \(s=\pm 1\) . 
\section{ {\,} Particles in a non-stationary external field}
As has been shown, in a stationary external field, particles of different charge signs with positive energies are described by independent Dirac equations. Let us now consider the case of a non-stationary electromagnetic field, assuming that the 4-vector potential is the sum of the stationary and non-stationary fields \(A_{\mu}(x)=A_{\mu}(\text{x})+\underline{A}_{\mu}(x)\) . We will assume that the average value of the variable field averaged over a large period of time is equal to zero 
\begin{equation} \label{97} \displaystyle\frac{1}{T}\int_{-T/2 }^{T/2}\underline{A}_{\mu}(x)dt
=0,\quad T\to \infty .
\end{equation} 
One particle of arbitrary sign in a non-stationary external field is also described by the Dirac equation \(Q_{\sigma}(x)=0\)  or \(\widetilde{\overline{Q}}_{\overline{\sigma}}(x)=0\) :
\begin{equation}\label{98}\displaystyle i\gamma^{\mu}\frac{\partial \psi_{\sigma}(x)}{\partial x^{\mu}}-\sigma\widetilde{e}A_{\mu}(\textbf{x})\gamma^{\mu}\psi_{\sigma}(x)-\mu\psi_{\sigma}(x)=\sigma\widetilde{e}\underline{A}_{\mu}(\textbf{x},t)\gamma^{\mu}\psi_{\sigma}(x). 
\end{equation}
If there are two particles of opposite signs in a non-stationary external field, then it follows from the equation \(Q_{\sigma}(x)+C^{\ast }\widetilde{\overline{Q}}_{\overline{\sigma}}(x)=0 \)  that their wave functions \(\psi_{\sigma}(x)\)  and \(\underline{\psi}_{\overline{\sigma}}(x)\equiv \widetilde{\overline{\psi}}_{\overline{\sigma}}(x)\)  are related: 
\begin{equation}\label{99}\displaystyle i\gamma^{\mu}\frac{\partial \psi_{\sigma}(x)}{\partial x^{\mu}}-\sigma\widetilde{e}A_{\mu}(x)\gamma^{\mu}\psi_{\sigma}(x)-\mu\psi_{\sigma}(x)=-\sigma\widetilde{e}C^{\ast }\underline{A}_{\mu}(\textbf{x},t)\widetilde{\gamma}^{\mu}\underline{\psi}_{\overline{\sigma}}(x),  \end{equation}
\begin{equation}\label{100}\displaystyle-i\widetilde{\gamma}^{\mu}\frac{\partial \underline{\psi}_{\overline{\sigma}}(x^{})}{\partial x^{\mu}}+\sigma\widetilde{e}A_{\mu}(x)\widetilde{\gamma}^{\mu}\underline{\psi}_{\overline{\sigma}}(x)-\mu\underline{\psi}_{\overline{\sigma}}(x)=\sigma\widetilde{e}\widetilde{C}\underline{A}_{\mu}(\textbf{x},t)\gamma^{\mu}\psi_{\sigma}(x).  
\end{equation} 
Thus, the states of two particles of different electric charges in an alternating external electromagnetic field are not independent. 
\section{Conclusion}

In its standard formulation, Dirac's theory describes particles with positive energy (electrons) and particles with negative energy, which are treated as particles of the opposite sign (positrons). It is assumed that the difficulty associated with the existence of negative energies can be overcome by moving to a quantum field description based on the apparatus of secondary quantization. However, as shown in \cite{Pl} and this work, even with a single-particle description, Dirac’s theory should be considered as a theory of particles of two types with the same mass, positive energies, and different charge signs. In this approach, the wave functions of particles admit a Born probabilistic interpretation and should be considered as probability amplitudes. In this paper, the description of electrons and positrons is based on the introduction of the concept of a charge quantum number, which characterizes the complex field of charged particles. It is natural to assume that a charged field of particles of any nature can be in two charge states and its interaction with the electromagnetic field should also be described using a charge quantum number. 

\end{document}